\documentclass{aa}  
\usepackage{lscape}
\usepackage{graphicx}

\usepackage{txfonts}
\usepackage{orcidlink}
\usepackage{textcomp, gensymb}
\titlerunning{Updated HIRES/Keck RV and Ca~{\sc ii}~H\&K catalog}

\begin{document}

   \title{An updated catalog of HIRES/Keck radial velocity measurements}

   \subtitle{Including Ca~{\sc ii}~H\&K measurements}

   \author{{Jerusalem T. Teklu} \orcidlink{0000-0002-4947-144X}
          \inst{1}\fnmsep\inst{2}
          \and 
          Volker Perdelwitz \orcidlink{0000-0002-6859-0882}
          \inst{3}
          \and
          R. Paul Butler \orcidlink{0000-0003-1305-3761}
          \inst{4}
          \and 
          Trifon Trifonov \orcidlink{0000-0002-0236-775X}
          \inst{2}\fnmsep\inst{5}
          \and
          Steven S. Vogt\inst{6}
          \and 
          Deepa Mukhija\inst{1}
          \and
          Lev Tal-Or \orcidlink{0000-0003-3757-1440}
          \inst{1}\fnmsep\inst{7}
          }

   \institute{Department of Physics, Ariel University, Ariel 40700, Israel\\
              \email{jerusalemt@msmail.ariel.ac.il}
         \and
        Universität Heidelberg, Landessternwarte, Zentrum für Astronomie, Königstuhl 12, 69117 Heidelberg, Germany
        \and 
        Department of Earth and Planetary Science, Weizmann Institute of Science, Herzl St 234, Rehovot, Israel 
        \and
        Earth and Planets Laboratory, Carnegie Science, 5241 Broad Branch Road NW, Washington, DC 20015, USA
        \and
        Department of Astronomy, Faculty of Physics, Sofia University ``St Kliment Ohridski'', 5 James Bourchier Blvd, 1164 Sofia, Bulgaria      
        \and
        UCO/Lick Observatory, Department of Astronomy and Astrophysics, University of California at Santa Cruz, Santa Cruz, CA 95064, USA
        \and
        Astrophysics, Geophysics and Space Science Research Center, Ariel University, Ariel 40700, Israel
             }

   \date{Received 4 April 2025 / Accepted 13 August 2025}

  \abstract
   {The first HIRES/Keck precision radial velocity (RV) catalog was released in 2017; it was followed by a second release in 2019, which incorporated corrections for small but significant systematic errors. The manifestation of stellar activity accompanied by systematic errors could affect the detection of exoplanets via the RV method.}
   {We expanded the HIRES catalog to March 2023 using publicly available spectra. Furthermore, we included the chromospheric emission line Ca~{\sc ii}~H\&K indicator ($R_{\mathrm{HK}}^\prime$), which is among the most prominent tracers of stellar activity.}
   {The precision RVs were obtained using an iodine gas absorption cell and corrected for minor systematic errors. $R_{\mathrm{HK}}^\prime$ measurements were derived by rectifying the observed spectra with PHOENIX synthetic spectra models in six narrow bands surrounding the H and K lines, then subtracting the photospheric contribution.}
   {We present an updated HIRES/Keck precision RV catalog featuring 78,920 RV measurements for 1,702 stars. High-quality $R_{\mathrm{HK}}^\prime$ measurements are provided for $\sim 40\%$ of the HIRES catalog. }
   {The updated catalog can help distinguish stellar activity effects from planetary signals in RV time series, thereby corroborating previously detected planetary candidates and aiding in the detection of new ones.}


   \maketitle

\section{Introduction} \label{sec:intro}
\citet{1952Obs....72..199S} proposed that if hot, Jupiter-mass planets orbit stars, they could potentially be detected even with 1950s-era technology using stellar radial velocity (RV) measurements and transit photometry surveys. Three decades later, \cite{1988ApJ...331..902C} carried out one of the earliest observational studies aimed at detecting exoplanets using the RV technique. At the same time, several researchers were monitoring the RVs of many stars in search of slight reflex orbital motions of the host star induced by a companion planet to detect exoplanet signatures \citep{1989Natur.339...38L, 1992PASP..104..270M, 1995Icar..116..359W}. The discovery of the hot Jupiter 51 Pegasi b by \cite{1995Natur.378..355M}, which is strikingly in a close-in orbit, marked the first confirmed detection of an exoplanet around a solar-type star via the RV method. This milestone provided compelling evidence for the existence of planetary systems beyond the Solar System, marking a pivotal advancement in modern astronomy.

As of June 2025, the RV method has led to the discovery of more than 1,100 out of the $\sim5,900$ confirmed exoplanets\footnote{Up to date list: \url{https://exoplanetarchive.ipac.caltech.edu/}}. Some of the most significant milestones in exoplanetary science include the occurrence-rate investigations that, among other findings, helped establish a clear distinction between planets and brown dwarfs \citep{2000PASPMarcy137M, 2006ApJGrether, 2010Sci...330..653H, 2011ASahlmann95S, 2015ARA&A..53..409W, 2023AJ....166..209M, 2024MNRAS.529.3958S, 2025arXiv250608078T}. A broader review of the occurrence rates, architectures, and demographic properties of planetary systems is given by \citet{2015ARAWinn409W}, and a recent census of exoplanet populations can be found in \citet{2018haex.bookE....D} and \citet{2024arXiv241000093B}.

Radial velocity measurements are derived from high-resolution spectroscopy and are used to detect minute periodic variations in stellar motion. The accuracy and precision of these measurements depend on the RV information content of the observed spectra \citep[e.g.,][]{1996ABaranneB, 2001ABouchy33B,  2020ApJSReinersR}, the stability of the wavelength calibration \citep{1992PASP..104..270M, 2016SPIE.9908E..6TJ, 2023JATIDebusD, 2024AReinersR}, and the precision of the instrumentation \citep{1996PASP..108..500B, 2016PASFischerF,  2017AJ....153..208B} and data reduction pipelines \citep{2005PASP..117..657W, 2014PASP..126..838W, 2020AJ....159..187P}. Early spectrographs could only achieve a precision of $\sim1$\,km\,s$^{-1}$ \citep[e.g.,][]{1967ApJ...148..465G}. Over time, the instrumental precision has significantly improved.

\citet{1973MNRAS.162..243G} proposed employing telluric absorption lines as a spectral reference to improve precision to $\sim0.01$\,km\,s$^{-1}$. Further advancements included the use of hydrogen fluoride absorption-cell spectrographs, which achieved a precision of $15$\,m\,s$^{-1}$ \citep{1979PASP...91..540C}, as well as iodine absorption cells, which refined precision to below $25$\,m\,s$^{-1}$ \citep{1992PASP..104..270M}. A breakthrough came with the development of the High Accuracy Radial velocity Planet Searcher (HARPS), a stable spectrograph that became the first instrument to achieve a precision of $1$\,m\,s$^{-1}$, significantly advancing exoplanet detection capabilities \citep{2002Msngr.110....9P, 2003Msngr.114...20M}. Building on this progress, efforts to further enhance RV precision continue today with next-generation instruments such as Echelle SPectrograph for Rocky Exoplanets and Stable Spectroscopic Observations (ESPRESSO), ESO's ultra-stable high-resolution spectrograph for the Very Large Telescope \citep[][]{2021A&A...645A..96P}. ESPRESSO has attained RV precision better than $25$\,cm\,s$^{-1}$ in a single night, demonstrating its potential to hit the instrument’s long-term stability target of $10$\,cm\,s$^{-1}$. These advancements are part of broader efforts, including extreme precision RV surveys \citep{2016PASFischerF}, that seek to detect Earth-mass analogs with velocity semi-amplitudes of approximately $10$\,cm\,s$^{-1}$ by improving RV measurement precision \citep[e.g.,][]{2021arXiv210714291C, 2023AJ....165..151N, 2024AJGupta29G}.

Among the first stable spectrographs that have significantly contributed to exoplanet studies is the High-Resolution Echelle Spectrograph \citep[HIRES;][]{1994SPIE.2198..362V}, which is mounted on the 10 m Keck I telescope atop Mauna Kea, Hawaii. In July 1996 it began to search for extrasolar planets around nearby F, G, K, and M dwarf stars. By utilizing iodine absorption lines between 5000 and 6200\,\AA \ for wavelength calibration \citep{1992PASP..104..270M} and covering a wavelength range of 3700 to 8000\,\AA , the instrument initially yielded a Doppler precision of $\sim3$\,m\,s$^{-1}$ \citep{1996PASP..108..500B}. HIRES has played a crucial role in several landmark discoveries of exoplanets, such as the confirmation of the first transiting extrasolar planet, HD\,209458\,b \citep{2000ApJ...529L..41H}, and the first Neptune-mass exoplanet, GJ 436b \citep{2004ApJ...617..580B}. In August 2004, the Lick-Carnegie Exoplanet Survey (LCES) team upgraded HIRES, improving its RV precision to $\sim1.5$ m s$^{-1}$ \citep{2006ApJ...646..505B}.

In a legacy paper published in 2017, LCES released all precise RVs of the HIRES/Keck dataset, encompassing 60,949 spectra for 1,624 stars observed over the course of more than two decades \citep[][hereafter B17]{2017AJ....153..208B}. Building on this, \citet[][hereafter TO19]{2019MNRAS.484L...8T} published an updated version of the catalog with 64,480 RVs for 1,699 stars, also correcting the RVs for small but significant systematic errors. The correction made use of RV-quiet stars (as defined in TO19 and described in Sect. \ref{subsec:rvnzp}) and included nightly zero-point (NZP) variations, as well as a discontinuous jump in August 2004 and an average intra-night drift. The use of RV standard stars to track and correct for systematic instrumental errors is currently a common practice. For example, \cite{2020ATrifonovT} used this approach for the HARPS/ESO RV dataset, and it is also used by the Calar Alto high-Resolution search for M dwarfs with Exoearths with Near-infrared and optical \'Echelle Spectrographs (CARMENES) RV survey \citep{2023A&A...670A.139R}.

In addition to the restrictions placed by instrumental noise, the magnetic activity of the host star can hinder the detection of exoplanets via the RV technique \citep[e.g.,][]{2001A&A...379..279Q, 2004ADesideraD, 2008A9Hulamo, 2014ApJ...796..132D, 2025arXivHatzes} and can even cause false-positive detections by mimicking an RV signal caused by a real planetary companion \citep[e.g.,][]{2022AJ....163..215S}. Stellar activity encompasses a combination of various phenomena such as pulsations, rotation, convection, and magnetic fields \citep[e.g.,][]{2012PhDT.......269D}. However, in this context, we primarily use the term to refer to the combined effect of surface inhomogeneity caused by convection, magnetic fields, and stellar rotation \citep{1984ApJ...279..763N, 1981A&A....96..345D, 2001A&A...374..733B, 2010A&A...512A..39M, 2024A&A...687A.303M}. Accurately determining the periodic modulation of stellar flux or the rotation period is essential to distinguishing between activity- and planet-induced RV variations \citep{2008MNRAS.386..516O, 2012MNRAS.419.3147A, 2015MNRAS.452.2269R}.

Various periodic and stochastic phenomena can induce RV modulations with a magnitude from a few up to a few hundred meters per second. These include solar-like activity cycles \citep{2010A&A...512A..39M}, granulation and supergranulation \citep{2011IAUS..276..527D, 2015A&A...583A.118M}, oscillations \citep{2011IAUS..276..527D}, flares \citep{2009A&A...498..853R}, and pulsations, especially in giant or subgiant stars \citep{1993ApJ...413..339H}. The RV modulations can occur over diverse timescales: years \citep{2011A&A...535A..55D}, days \citep{2012PhDT.......269D}, hours \citep{2008PASP..120.1043F, 2009A&A...498..853R}, and minutes \citep{2019AJ....157..163C}.

The rotation periods of main-sequence stars range from a fraction of a day to about 200 days \citep{2014ApJS..211...24M} and generally are determined using multiple activity indicators. If accurately used, each of these can provide information on the rotation period \citep{2022A&A...663A..68S, 2024A&A...684A...9S} or cycle \citep{1998ApJ...498L..51B, 2025MNRAS.540..668C}. These indicators can be categorized as photospheric or chromospheric, based on the origin of the spectral lines involved in their calculation. The prominent photospheric indices are the cross-correlation function (CCF) moments, including RVs \citep[e.g.,][]{1996ABaranneB} and the chromatic index \citep{2018A&A...609A..12Z}. The best-known indicators of chromospheric activity include emission lines, mainly from plages and filaments, such as the Ca~{\sc ii}~H\&K doublet \citep{1968ApJ...153..221W, 1984ApJ...279..763N, 2021A&A...652A.116P}---which comprises the $K$ line at $3933.660$\,{\AA} and the $H$ line at $3968.47$\,{\AA}---the Ca~{\sc ii} infrared-triplet \citep{2017A&A...607A..87M}, and the $H_\alpha$ line \citep{1986ApJS...60..551F}. 

Various methods for extracting chromospheric flux excess---particularly that of the most prominent indicator, the Ca~{\sc ii}~H\&K index---have been well established. Among the most widely used techniques is the S index, which directly measures the emission in the Ca~{\sc ii}~H\&K lines relative to adjacent continuum regions \citep{1968ApJ...153..221W, 1995ApJ...438..269B}. \cite{1982A&A...107...31M} proposed a method for converting the S index into absolute flux units utilizing the stellar (B-V) color index. \cite{1984ApJ...279..763N} further refined this approach by developing a procedure for photospheric contributions to the total H\&K flux and normalizing the Mount Wilson S index to the star's bolometric luminosity. This process resulted in another index, $R_{\mathrm{HK}}^\prime$, which is less prone to bias by stellar luminosity or temperature and enabled more direct comparisons of chromospheric activity between stars of different types. An alternative method, spectral subtraction, involves removing the spectrum of a photospherically inactive star (or a model atmosphere) from the spectrum of an active star \citep{1985ApJ...295..162B, 1995A&AS..114..287M, 2011A&A...531A...8J, 2016A&A...586A..14H} and attributing the residual flux to chromospheric activity. This technique can theoretically be applied to any spectral line, not only Ca~{\sc ii}~H\&K. While studies have also calibrated the S index and its conversion to the $R_{\mathrm{HK}}^\prime$ index \citep{2015MNRASMascareS, 2017A&A...602A..88A}, \citet{2021A&A...652A.116P} introduced a method for directly measuring the relative chromospheric flux in the H\&K lines, rectifying spectra using PHOENIX stellar atmospheres and synthetic spectra \citep{2013A&A...553A...6H} and narrow passbands centered around the emission line cores. We used this method to derive $R_{\mathrm{HK}}^\prime$ for the entire public HIRES dataset. A similar update has also been applied to the HARPS dataset \cite{2024A&A...683A.125P}.

This work is structured as follows: Section \ref{sec: rvbank} details the extension of the HIRES RV catalog and the extraction of $R_{\mathrm{HK}}^\prime$. Section \ref{sec: result} presents the results and compares them with previous studies, and Sect. \ref{sec: summary} provides a summary and conclusions. 

\section{Updating the HIRES catalog} \label{sec: rvbank}
This paper presents the updated HIRES RV catalog as of March 2023, incorporating an addition of 14,440 spectra and three newly included stars beyond those listed in TO19, bringing the total number of stars to 1,702. This update also contains corrections for minor systematic errors in RV measurements, following the approach in TO19, and the addition of $R_{\mathrm{HK}}^\prime$ from individual spectra. The determination of $R_{\mathrm{HK}}^\prime$ was carried out for the entire catalog by using the method of \citet{2021A&A...652A.116P}. However, successful measurements were obtained only for stars that met both of the following criteria: sufficient signal-to-noise ratio in both of H\&K lines, and spectra with no issues related to observation or calibration. Eventually, $R_{\mathrm{HK}}^\prime$ measurements are provided for $\sim 40\%$ of the catalog.

\subsection{Stellar sample and parameters} \label{subsec:stellarpara}
\begin{figure}
\resizebox{\hsize}{!}{\includegraphics[width=0.48\textwidth]{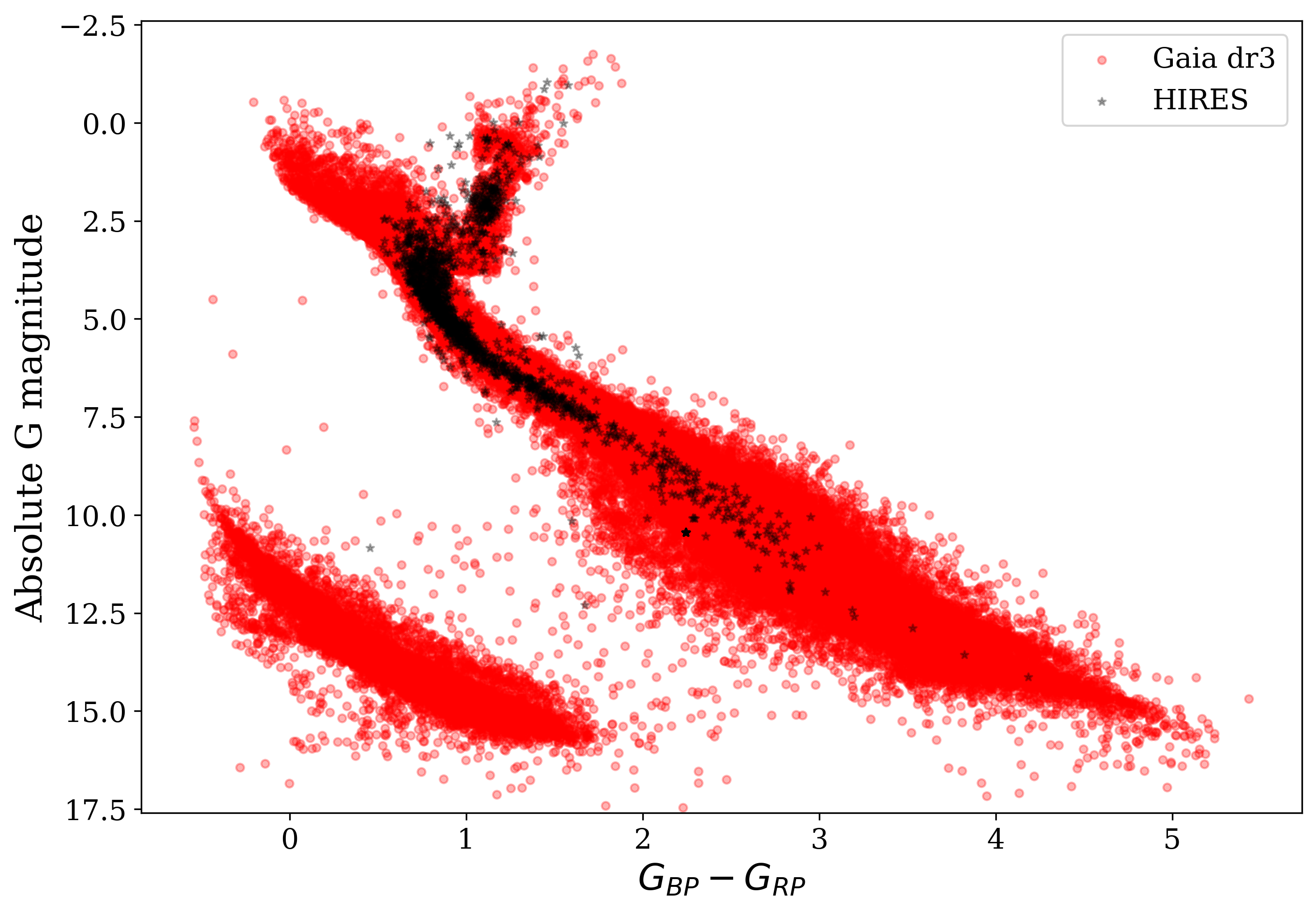}}
\caption{\label{fig: 1gaia}Distribution of HIRES target stars (black) compared to that of \textit{Gaia} DR3 stars (red).} 
\end{figure}

Among the 1,702 HIRES target stars, roughly $74\%$ are classified as main-sequence stars of spectral types F, G, K, or M, approximately $9\%$ as subgiants, around $10\%$ as giants, and the remaining $\sim 7\%$ lack measurements of $\log g$. The distribution of HIRES target stars as seen by \textit{Gaia} Data Release 3 \citep[DR3;][]{2023AGaia39G} is shown in Fig. \ref{fig: 1gaia}. For this plot, we excluded stars with potentially unreliable \textit{Gaia} data—specifically, those showing signs of uncertain motion (astrometric excess noise $>$ 1\,mas) or suspicious brightness and/or color measurements (e.g., very high BP-RP color excess). 

The stellar input parameters—effective temperature ($T_{\rm eff}$), surface gravity ($\log g$), metallicity ($[M/H]$), projected rotational velocity ($v \sin i$), and the spectral resolution ($\mathcal{R}$) of the HIRES instrument—are utilized to estimate theoretical flux values that closely match the observed stellar properties within the grid of PHOENIX synthetic spectra. When $v \sin i$ values were not available, we used the line-broadening values measured with a rotational kernel (vbroad) as an alternative. For F, G, and K-type stars, vbroad and $v \sin i$ generally agree, except in cases where $v \sin i < 10$ km\,s$^{-1}$, where vbroad can be biased by up to 5 km\,s$^{-1}$, as reported by \citet{2023Amat8F}.

\begin{figure*}
\centering
\includegraphics[width=1\textwidth]{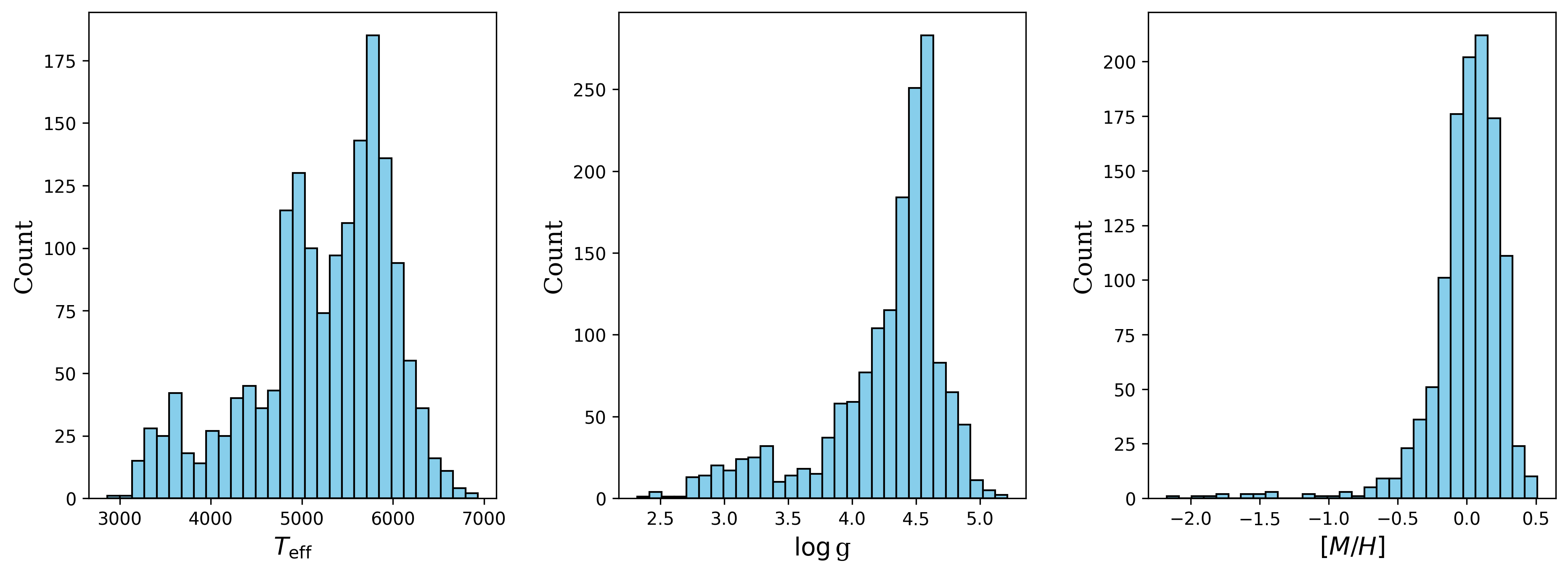}  
\caption{\label{fig: disstella} Distribution of the overall stellar parameters for the entire catalog: $T_{\rm eff}$ (left), $\log g$ (center), and $[M/H]$ (right).}    
\end{figure*}

Furthermore, the absolute RV is required to transform the observed spectrum into the rest frame. All stellar parameters along with their uncertainties were obtained from DR3 \citep{2023AGaia39G} and the TESS Input Catalog \citep[][]{2019AJ....158..138S} via VizieR \citep{2000A&AS..143...23O}. Some of the RV data, in addition to \textit{Gaia}, were obtained from \citet{2007AN....328..889K, 2000A&AS..143....9W,
2015ApJ...802L..10T}; and \citet{2014MNRAS.444.3517M}. \autoref{table: 1param} presents a representative sample of the stellar parameter dataset. In addition, Fig. \ref{fig: disstella} shows the $T_{\rm eff}$, $\log g$, and $[M/H]$ distributions for the full stellar sample. Indeed, the majority of stars are Solar-metallicity FGK dwarfs.

\subsection{Derivation of RVs}
Doppler measurements were made using the iodine technique pioneered by \cite{1996PASP..108..500B}. An evacuated 4-inch-long Pyrex absorption cell is filled with iodine vapor at a temperature of $\approx$ 37\,\degree C, corresponding to an atmospheric pressure of 0.003 atm. The cell is then operated at a temperature
of 50\,\degree C, so that all iodine is in the vapor phase, and the column density of iodine vapor is invariant. This was experimentally shown by \citep{2018MNRAS.479..768P}. The absorption cell is mounted directly in front of the entrance slit of the HIRES spectrometer, superimposing iodine lines on the stellar spectrum. The iodine lines provide both an absolute wavelength scale and a proxy for the shape of the spectrometer point spread function (PSF), which is determined anew for each spectrum and at each wavelength. All time-dependent wavelength-scale changes and spectrometer PSF asymmetries are included in the Doppler analysis at each wavelength.

The iodine cell also ensures the stability of wavelength calibration and PSF information on timescales up to decades, enabling the search for giant planets in orbits of similar size to those in our Solar System. Indeed, the Keck iodine cell has not been altered since the beginning of data acquisition in 1996, preserving the integrity (velocity zero-point and scale) of the velocity measurements, despite any changes to the optics or detector of the HIRES spectrometer. The spectrometer is operated with a resolution of R$\approx 70000$ and a wavelength range of 3700--6200\,\AA\,\citep{1994SPIE.2198..362V}, although only the region 5000--6000\,\AA\,(i.e., the region containing iodine lines) was used in the Doppler analysis. The Doppler shifts from the spectra are determined with the spectral synthesis technique described by \citet{1996PASP..108..500B}.

\begin{figure} 
\resizebox{\hsize}{!}{\includegraphics[width=0.44\textwidth]{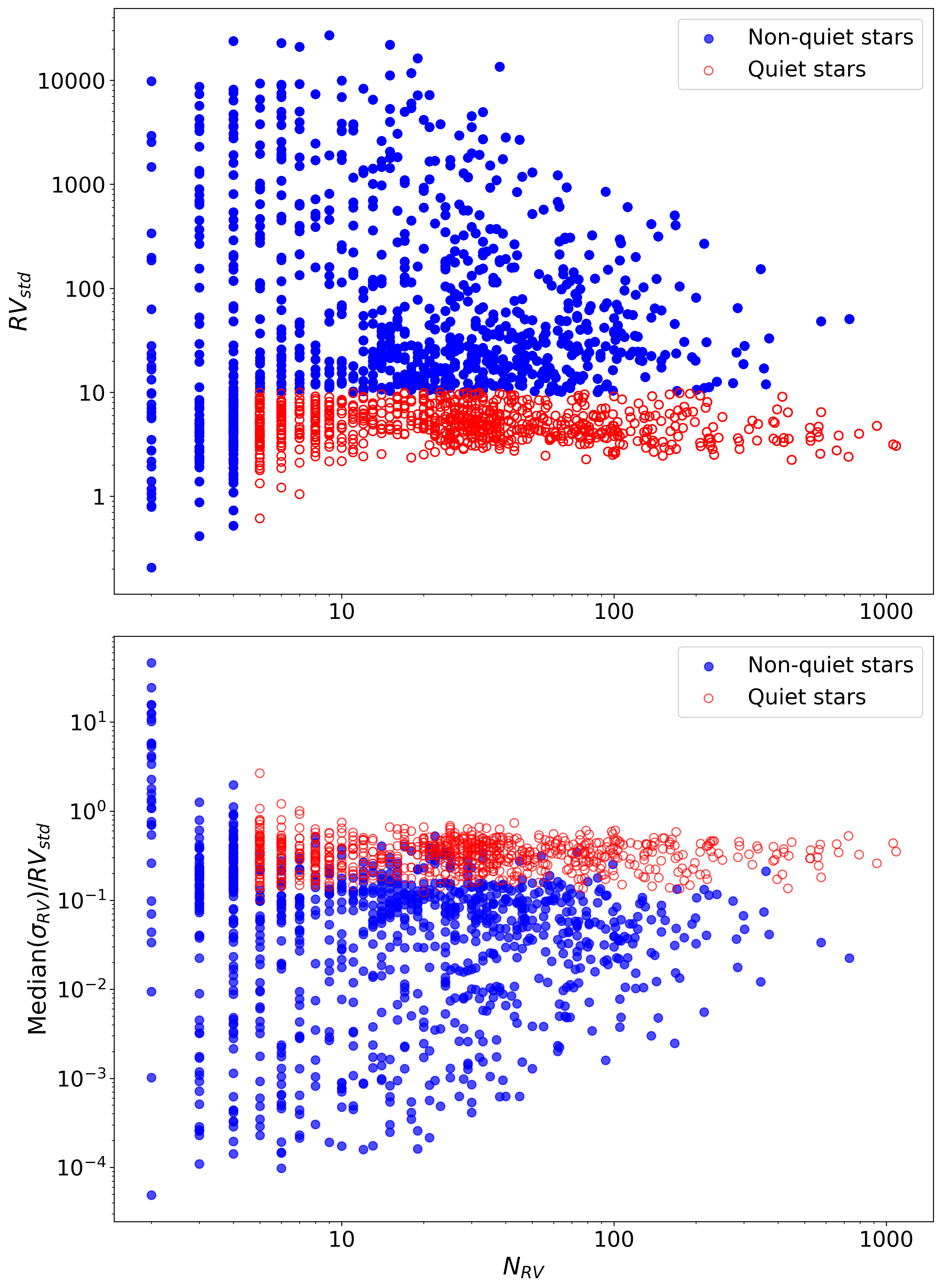}}
\caption{\label{fig: 3rvdis}Representation of RV measurements color-coded by RV-quiet threshold criteria: quiet in red and non-quiet in blue. Top: Standard deviation of the (NZP-corrected) RVs as a function of the number of RV observations ($N_{RV}$). Bottom: Ratio between the median RV uncertainty and the standard deviation of the corrected RVs, providing a measure of how well the RV precision accounts for the observed scatter.}
\end{figure}

The illustration of whether the observed variations in a star’s RV are due to real astrophysical signals, such as planets, stellar activity, or binaries, or merely due to measurement noise is presented in Fig. \ref{fig: 3rvdis}. As expected, RV-quiet stars, defined as those with more than five RV measurements and an RV standard deviation ($RV_{\rm std}$) below $10$\,m\,s$^{-1}$, converge to $RV_{\rm std}\sim 3$\,m\,s$^{-1}$ as the number of observations increases. Naturally, non-quiet stars generally show higher RV variability. This trend is visible in the top panel of the figure.

At $N_{\rm RV}\leq7$, both quiet and non-quiet stars display a large spread in $RV_{\rm std}$, likely due to poor or short period sampling. This emphasizes the need for caution when interpreting variability in poorly sampled stars. Beyond about $7$ RV measurements, the lower envelope of the distribution (i.e., the minimum $RV_{\rm std}$ across $N_{RV}$) represents the external HIRES noise floor of $\sim 2$\,m\,s$^{-1}$.

The bottom panel of Fig. \ref{fig: 3rvdis} shows, per star, the ratio between the median internal RV uncertainty and the observed external $RV_{\rm std}$. As $N_{RV}$ increases, this ratio stabilizes. Quiet stars tend to converge toward a ratio of $\sim \frac{1}{3}$ with an upper envelope of $\sim \frac{1}{2}$. This indicates that internal measurement errors only account for up to half of the observed variability. Since instrumental errors are $\leq 1$\,m\,s$^{-1}$ (Sect. \ref{subsec:rvnzp}), the residual variability is probably driven by stellar activity and orbital motion. In contrast, for non-quiet stars the ratio is typically below $0.2$, reflecting intrinsic astrophysical RV variability.

In addition to the precise relative RVs, we derived absolute RVs by computing the CCF for each of the eight blue-most echelle orders spectra in the HIRES dataset using the Python package PyAstronomy\footnote{\url{https://pyastronomy.readthedocs.io/en/latest/pyaslDoc/aslDoc/crosscorr.html}}. We used PHOENIX synthetic spectra as templates, selecting the appropriate model for each star based on its stellar parameters: $T_{\rm eff}$, $\log g$ and $[M/H]$. The template's wavelength axis was adjusted using the proper Doppler shift to account for each RV shift. The shifted template was then linearly interpolated onto the wavelength grid of the observed spectrum to compute the CCF.

We then used the catalog RVs for an initial RV estimate, constraining the search range to $\pm 150$\,km\,s$^{-1}$ around the initial RV to refine the RV determination. This approach optimizes computational efficiency and minimizes the risk of missing the correct CCF peak due to uncertainties in catalog values. Additionally, we excluded the first and last 20 data points from each spectrum to enhance measurement reliability and mitigate contamination from spectral edges before cross-correlation. This precaution helps reduce the influence of noise and instrumental artifacts often present at the spectrum boundaries. We used the per-order RV at the CCF maximum and calculated the difference from the catalog RV, flagging cases where it exceeds $3$\,km\,s$^{-1}$, as discussed in Sect. \ref{sec: extract}.

\begin{figure*} 
\centering
\includegraphics[width=\textwidth]{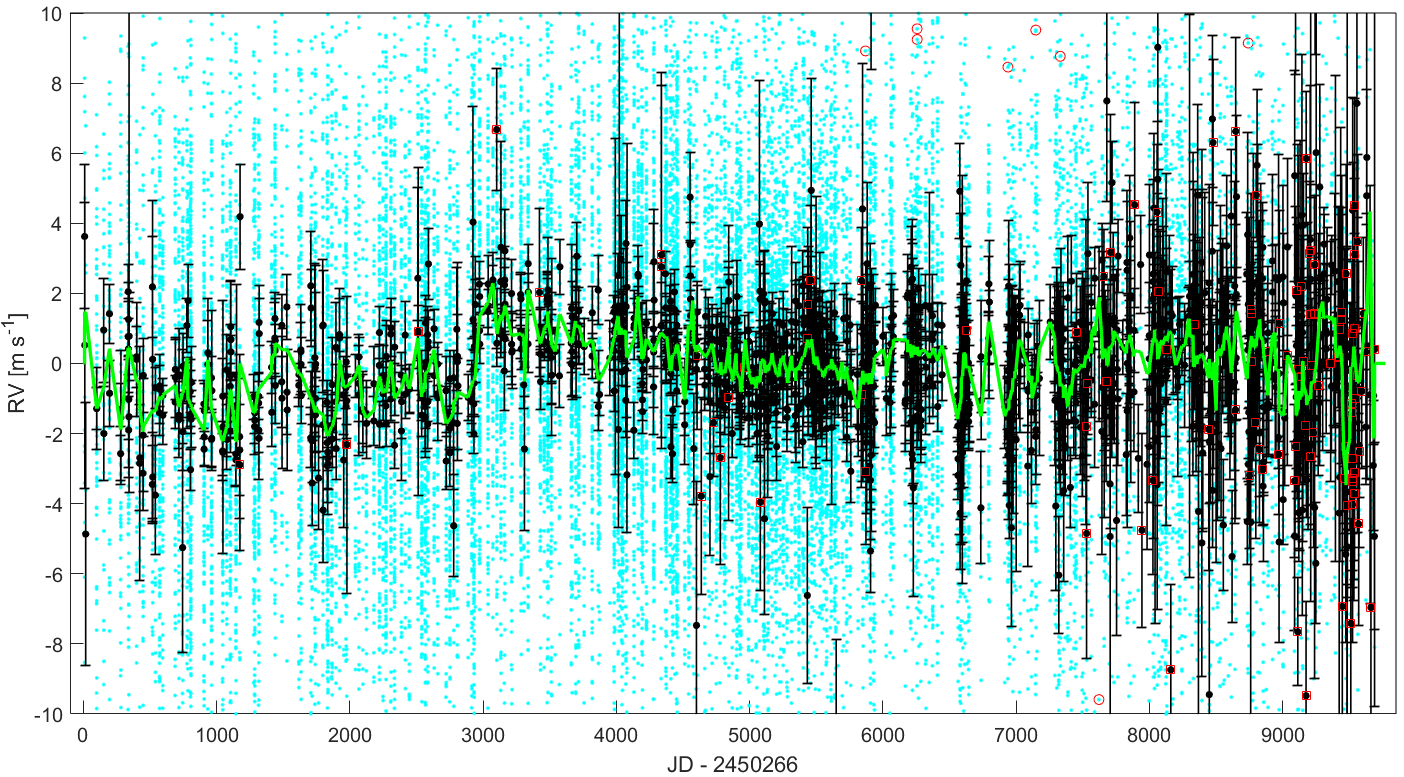}
\caption{\label{fig: 2nzp}Systematic variability of the HIRES instrument. Individual RVs of RV-quiet stars are denoted in cyan, and nightly weighted average RVs (NZPs) in black. NZPs enclosed in red boxes represent those derived from fewer than three RV measurements, while the green line indicates the 50-day moving average NZP model.}
\end{figure*}

\subsection{Systematic variations in the relative HIRES RVs} \label{subsec:rvnzp}
The relative HIRES RVs were corrected for NZP variations and an average intra-night drift (from evening to morning) using the method described in TO19. For HIRES, these variations can introduce shifts of up to $1.0$\,m\,s$^{-1}$ rms. Figure \ref{fig: 2nzp} displays the individual RVs of RV-quiet stars, used to compute the NZPs, along with the resulting NZPs. The high noise in NZPs observed in recent data is likely a consequence of fewer observations per night, as marked by the red boxes, which denote NZPs derived from fewer than three RV measurements.

\begin{table*}
\centering
\footnotesize
\setlength\tabcolsep{2.5pt}
\caption{Main stellar parameters of the HIRES sample (extract). }
\label{table: 1param}
\begin{tabular}{c c c c c c c c c c c c c c c c c c}
\hline\hline
Name & Ra & Dec & RV & e\_RV & Ref. & Teff & e\_Teff & Ref. & $\log g$ & e$\_\log g$ & Ref. & [M/H] & e$\_[M/H]$ & Ref. & $v \sin i$ & e$\_v \sin i$ & Ref.\\
\hline
& [deg] & [deg] & [km\,s$^{-1}$] & [km\,s$^{-1}$] & & [K] & [K] & &  &  & & [dex] & [dex] & & [km\,s$^{-1}$] & [km\,s$^{-1}$] & \\
\hline
HD\,7924 & 20.4963 & 76.7102 & -22.7 & 0.15 & 1 & 5136.0 & 143.3 & 2 & 4.58 & 0.093 & 2 & -0.14 & 0.1 & 2 & \ldots & \ldots & \ldots \\
HD\,219134  & 348.3207 & 57.1684 & -18.64 & 0.19 & 1 & 4884.3 & 142.8 & 2 & 4.59 & 0.099 & 2 & 0.095 & 0.021 & 2 & \ldots & \ldots & \ldots \\
HD\,156668 & 259.4187 & 29.2272 & -44.57 & 0.17 & 1 & 4849.9 & 103.0 & 2 & 4.574 & 0.077 & 2 & -0.051 & 0.047 & 2 & 7.780 & 0.848 & 1 \\
HD\,185144 & 293.0899 & 69.6611 & 26.58 & 0.16 & 1 & 5242.0 & 138.7 & 2 & 4.59 & 0.090 & 2 & -0.2 & 0.1 & 2 & \ldots & \ldots & \ldots \\
HD\,42618 & 93.0023 & 6.7830 & -53.52 & 0.14 & 1 & 5755.0 & 115.0 & 2 & 4.47 & 0.073 & 2 & -0.11 & 0.1 & 2 & 5.406 & 0.861 & 1 \\
\hline
\end{tabular}
\tablebib{(1)~\citet{2023AGaia39G};  (2)\citet{2019AJ....158..138S}.}
\tablefoot{ The full table can be accessed at the CDS. Coordinates are in J2000 International Celestial Reference System (ICRF) frame.}
\end{table*}

\begin{figure}
\resizebox{\hsize}{!}{
\includegraphics[width=0.5\textwidth]{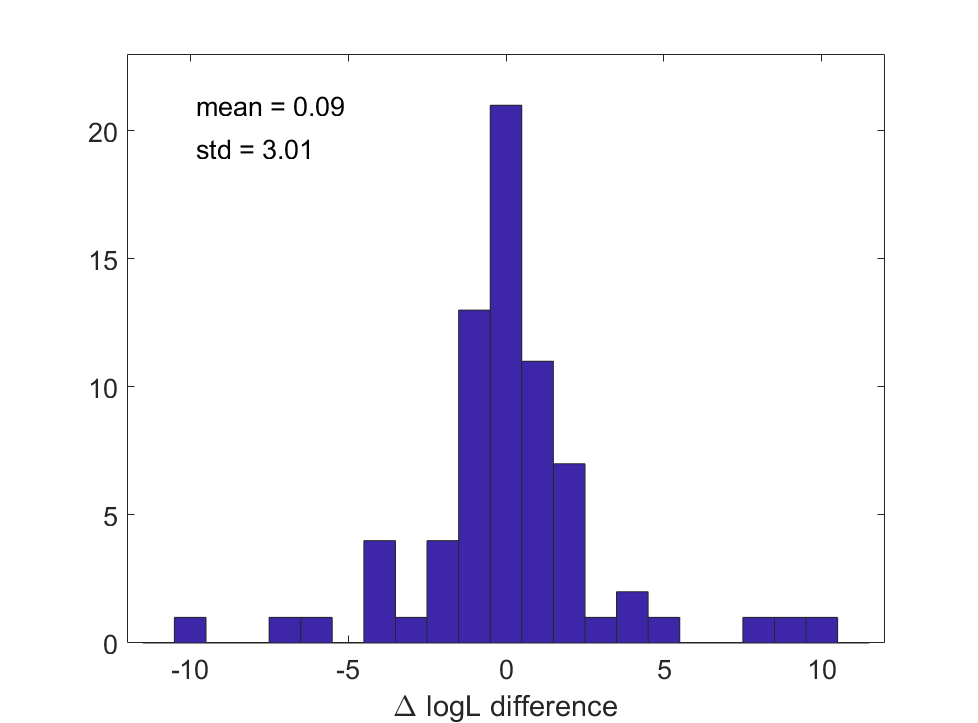}}
\caption{\label{fig: 5logl}Distribution of the $\Delta\log\mathcal{L}$ difference of fitting the HIRES RV data of 71 stars known to host a single massive planet with a Keplerian model. We repeated the fit twice, once with the original (uncorrected) RVs and once with the corrected RVs, and calculated the difference in log-likelihood improvement ($\Delta\log\mathcal{L}$) with respect to a weighted-mean model of a constant RV star. The list of planets and corresponding $\Delta\log\mathcal{L}$ values is given in \autoref{table:appendi}.}
\end{figure}

To evaluate the effectiveness of our NZP correction approach, we compared the RV data likelihood before and after applying the correction, using a sample of confirmed planetary signals\footnote{\url{https://exoplanetarchive.ipac.caltech.edu/}}. Specifically, we modeled the RVs of 71 stars known to host a single massive planet using a Keplerian fit. An additional linear trend was added to the model where necessary. Each star was fitted twice: once using the original (uncorrected) RVs and once using the NZP-corrected RVs. For each star, we computed the difference in log-likelihood improvement relative to a weighted-mean model that assumes a constant RV. The distribution of these differences, denoted by $\Delta\log\mathcal{L}$, is presented in Fig. \ref{fig: 5logl}. It is approximately normal, with a mean of 0.09 and a standard deviation of 3.01. This indicates that, on average, the NZP correction only marginally improves the HIRES RVs.

This result is consistent with expectations. The long-term NZP stability of HIRES is better than $1$\,m\,s$^{-1}$, aside from a known $\sim 1.5$\,m\,s$^{-1}$ offset in 2004. In comparison, the median RV uncertainty is $1.5$\,m\,s$^{-1}$, and the observed RV scatter is significantly larger. The typical RV scatter of $>3$\,m\,s$^{-1}$ is primarily driven by stellar activity and orbital dynamics, rather than by internal RV uncertainties and instrumental errors. Nevertheless, to maintain consistency with previous catalog versions, and to allow flexibility for future users, we provide both the corrected and uncorrected RV datasets, enabling independent decisions on whether or not to use the correction.

\subsection{Extraction of \texorpdfstring{$\text{R}'_\text{HK}$}{R'HK}} \label{sec: extract} 
In order to extract \texorpdfstring{$\text{R}'_\text{HK}$}{R'HK}, we retrieved the entire archive\footnote{\url{https://koa.ipac.caltech.edu/cgi-bin/KOA/nph-KOAlogin?more}} of pipeline-reduced HIRES spectra for the wavelength range of $\sim 3200 - 4500$\,{\AA} in single-order format. 
The $R_{\mathrm{HK}}^\prime$ value was then determined by employing the method introduced by \cite{2021A&A...652A.116P}, with a modification to the $k_1$ band-pass around the K line to minimize the impact of absorption lines, as described in \cite{2024A&A...683A.125P}. This approach enabled us to calculate the relative flux excess in the chromosphere ($R_{\mathrm{HK}}^\prime$) by rectifying the observed spectrum with synthetic PHOENIX spectra using six narrow bands centered on the H and K lines and subsequently subtracting the photospheric flux.

During this process, we identified several quality issues within individual spectra, and after investigating potential causes, we flagged each spectrum accordingly. We defined four custom flags based on our data analysis and one internal HIRES flag. Two flags pertain to flux calibration issues: flux values below zero and regions exhibiting plateau-like flat characteristics. Another flag identifies folded spectra with negative values of the wavelength first derivative, and the final flag marks cases where the difference between the CCF RV derived by us and the catalog RV exceeds $3$\,km\,s$^{-1}$. To ensure the integrity of our derivation process, we utilized a subset of the data, referred to as the ``clean sample,'' that comprises approximately $ 21\%$ of the entire sample of $R_{\mathrm{HK}}^\prime$ measurements, which contains no flagged entries. 

\section{Results and discussion}
\label{sec: result}

Radial velocities with NZP correction and $R_{\mathrm{HK}}^\prime$ measurements were derived from approximately 29 years of high-precision spectroscopic observations. We obtained 78,920 RV measurements for 1,702 target stars, incorporating an additional 14,440 spectra and three new stars compared to TO19.

Utilizing the methodology outlined in Sect. \ref{sec: extract}, $R_{\mathrm{HK}}^\prime$ values were extracted for 31,218 spectra, representing $40\%$ of the total of 78,920 spectra. The availability of $R_{\mathrm{HK}}^\prime$ measurements was limited by data quality, with the highest confidence subset, called the clean sample, comprising 1,285 stars out of the entire catalog of 1,702 stars. In the HIRES sample, this subset represents the most reliable data for analyzing stellar activity and its impact on RV measurements. \autoref{table: 2rvlist1} presents a representative excerpt of the final catalog, showing the derived parameters and their assigned flags. 

Beyond the catalog description, in what follows, we provide an overview of a couple of statistical features of our RV and $R_{\mathrm{HK}}^\prime$ measurements in \autoref{table: 3derista}, as well as a few representative examples illustrated in Fig. \ref{fig: examples}.

\begin{table*}
\centering
\small 
\setlength\tabcolsep{5pt}
\caption{Statistical overview of the derived parameters (extract).} 
\label{table: 3derista}
\begin{tabular}{c c c c c c c c c c c}
\hline\hline
Name & Ra & Dec & Time\_span & nRV & $RVC_{\rm std}$ & $e\_RVC$\_Med & $R_{\mathrm{HK}}^\prime$\_Med & $dR_{\mathrm{HK}}^\prime$\_Med & S index\_Med & H-index\_Med \\
\hline
& [deg] & [deg] & [d] &  & [m\,s$^{-1}$] & [m\,s$^{-1}$] &  &  &  &\\
\hline
HD\,7924 & 20.4963 & 76.7102 & 7701.93 & 923 &  4.76 & 1.24 & 8.24$ \times 10^{-6} $ & 1.47$ \times 10^{-6} $ & 0.21 & 0.03\\
HD\,219134 & 348.3207 & 57.1684 & 9557.96 & 576 & 6.45 & 1.16 & 1.38$ \times 10^{-5} $ & 1.35$ \times 10^{-6} $ & 0.26 &  0.04 \\
HD\,156668 & 259.4187 & 29.2272 & 6878.96 & 565 & 3.74 & 1.38 & 1.30$ \times 10^{-5} $ & 1.06$ \times 10^{-6} $ & 0.25 & 0.04  \\
HD\,185144 & 293.0899 & 69.6611 & 9226.69 & 1086 & 3.06 & 1.07 & 6.27$ \times 10^{-6} $& 1.24$ \times 10^{-6} $ & 0.19 & 0.03\\
HD\,42618 & 93.0023 & 6.7830 & 9086.02 & 668 & 3.84 & 1.32 & -4.08$ \times 10^{-6} $ & 6.39$ \times 10^{-7} $ & 0.15 & 0.03\\
\hline
\end{tabular}
\tablefoot{ Time\_span is the total time of observation, and nRVis the number of RV measurements. All parameters with Med indicate the median values of certain parameters described in \autoref{table: 2rvlist1}. The complete table, appended to \autoref{table: 1param}, can be accessed at the CDS.}
\end{table*}

\subsection{RV scatter and rotational velocity}

The typical RV uncertainty is $1.4$\,m\,s$^{-1}$, which sets an instrumental limit on the most minor detectable RV variations. Furthermore, the uncertainty of the NZP correction is characterized by a median value of $0.41$\,m\,s$^{-1}$, indicating that the corrections are generally precise and most measurements exhibit a relatively low uncertainty. The standard deviation of the corrections, $0.86$\,m\,s$^{-1}$, being more than twice the median uncertainty points to a potential improvement in the accuracy of the RVs. 

However, Fig. \ref{fig: 5logl} shows that, based on a sample of single-planet host stars, HIRES RVs are only marginally improved by the correction. Given that the typical correction is much smaller than the typical uncertainty, this result is not surprising. Importantly, this is not the case for other precision RV instruments for which the method was applied. The NZP variability for HARPS is $\sim1.5$\,m\,s$^{-1}$ \citep{2020ATrifonovT} and for CARMENES it is is $\sim2.3$\,m\,s$^{-1}$ \citep{2023A&A...670A.139R}. At the same time, the typical internal RV uncertainties for HARPS are $\sim1.0$\,m\,s$^{-1}$, and for CARMENES they are $\sim1.5$\,m\,s$^{-1}$. Therefore, while for HIRES, the correction is typically smaller than the uncertainty, for HARPS and CARMENES, it is the other way around.

\begin{figure} 
\resizebox{\hsize}{!}{\includegraphics[width=0.5\textwidth]{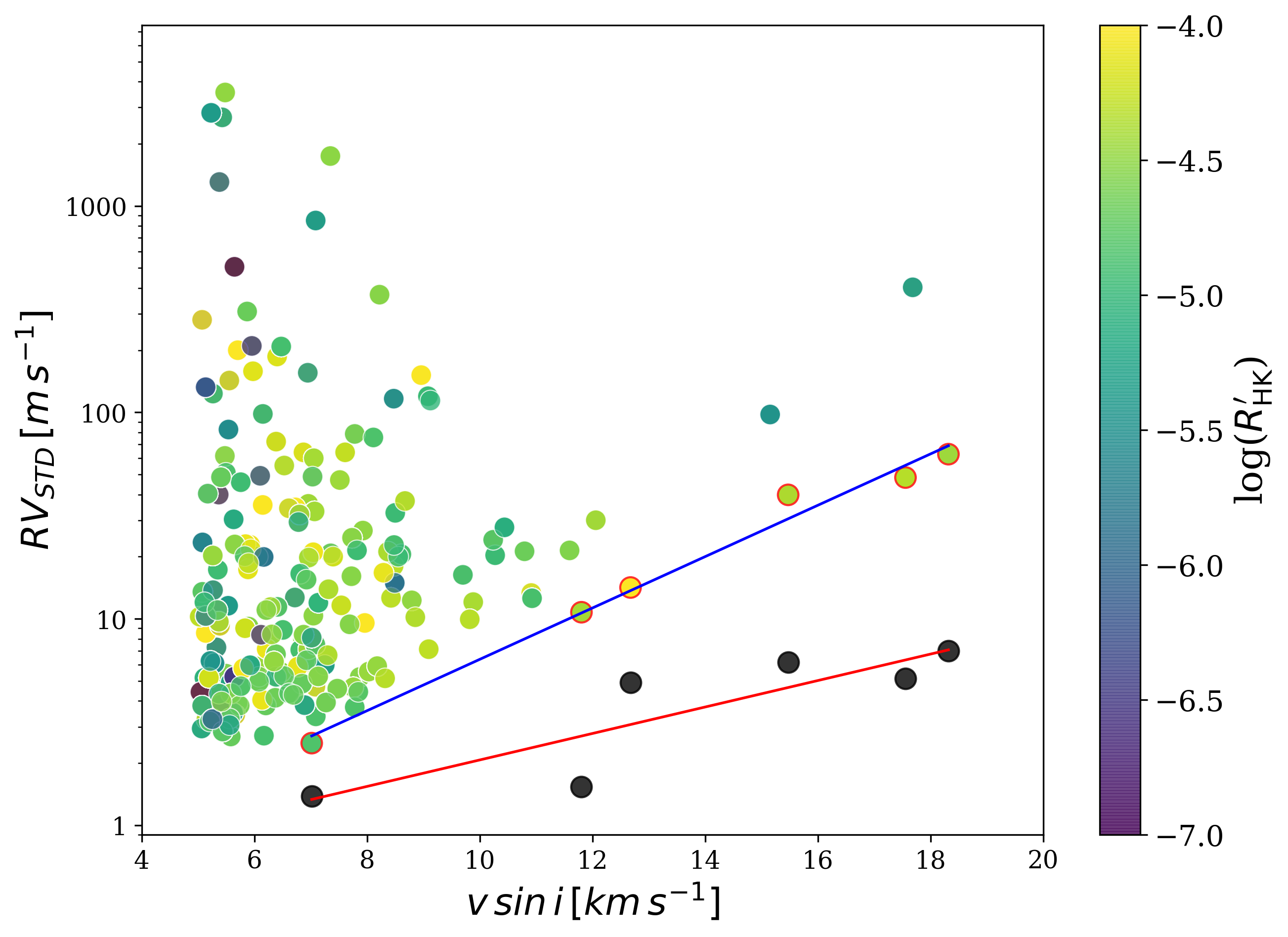}}
\caption{\label{fig: rvsini} $RV_{\rm std}$ as a function of the projected rotational velocity. The blue and red lines are two linear regression fits: blue for the lower-envelope $RV_{\rm std}$ values of the six stars whose RVs are dominated by stellar activity rather than by orbital motion (from left to right: HD202751, HD214749, HD210302, HD21847, HD377, and HD25998), and red for the median RV uncertainties of the same stars. Data points are color-coded by $\log(R_{\mathrm{HK}}^\prime)$ to indicate chromospheric activity levels. Since \textit{Gaia} has known problems with correctly measuring the $v\,\rm sin\,i$ of cool dwarfs ($T_{\rm eff} < 4300$\,K), we removed cool dwarfs from the plot and only make line fits in the $v\,\rm sin\,i > 7$ km\,s$^{-1}$ region.} 
\end{figure}

In addition to instrumental limitations, stellar rotation and surface inhomogeneities, such as starspots, introduce variability in the observed RV measurements. In Fig. \ref{fig: rvsini} we present the relationship between the RV standard deviation per star ($RV_{\rm std}$) and the projected rotational velocity ($v\,\rm sin\,i$) for our selected clean sample. The $RV_{\rm std}$ values displayed here have been derived from NZP-corrected RVs, while the $v \sin\,i$ values are sourced from catalogs, predominantly \textit{Gaia}. 

We excluded cool dwarfs ($T_{\mathrm{eff}} < 4300$ K) because, as noted in \cite{2023ABlanco29R}, \textit{Gaia}-based stellar parameters are not well constrained for such stars. In particular, the $v \sin\,i$ values of slow-rotating stars may be unreliable. The challenge in estimating their stellar atmospheric parameters may stem from the complexity of their atmospheres and the effects of convective mixing \citep{2024ApJSQu2Q}. Furthermore, $v\,\rm sin\,i$ values smaller than $10$\,km\,s$^{-1}$ are likely upward biased \citep{2023Amat8F}, with the bias growing for smaller $v sin\,i$ values. Therefore, we only focused on the RV noise floor in the $v \sin\,i > 7$km\,s$^{-1}$ region.

Our analysis reveals that, in HIRES, the RV variability of stars exhibits a lower limit of approximately $2.5$\,m\,s$^{-1}$ for the slow rotators. For stars with $v\,\rm sin\,i < 7$\,km\,s$^{-1}$, the leading sources of RV variability would rather be internal uncertainty and orbital motion than stellar activity. A similar observation was made by \cite{2024A&A...692A.138R}, who reported a jitter floor of $2$\,m\,s$^{-1}$ for M-dwarf stars with low rotation velocities, consistent with the noise floor of CARMENES \citep{2018SPIE10702E..0WQ, 2023A&A...670A.139R}. From a different point of view, based on HARPS measurements, \cite{2017MNRAS.468.4772S} demonstrated that chromospheric activity indicators, such as $\log(R_{\mathrm{HK}}^\prime)$, correlate with RV variability down to $RV_{\rm std}$ values of $\sim1$\,m\,s$^{-1}$ for M dwarfs, and below that for G and K dwarf stars. All three works suggest that precision RV surveys with instruments having a noise floor limit of $1-2$\,m\,s$^{-1}$, such as HIRES, CARMENES, and HARPS, may include stars whose RV variability is not dominated by stellar activity. These would predominantly be slow rotators.

For faster rotating stars, $v\,\rm sin\,i$ appears to be a reliable parameter for assessing the attainable RV precision for a given star. The trend, derived from the linear fit to the lower envelope of $RV_{\rm std}$ values (blue line in Fig. \ref{fig: rvsini}), provides an empirical framework for predicting the expected minimum level of RV jitter based on stellar rotation. To verify that stellar activity, rather than insufficient RV information content, dominates the RV variability of fast rotating stars, we show, for the six stars that outline the lower envelope of $RV_{\rm std}$, the median RV uncertainty per star, as well as a line fit to these values (red line in Fig. \ref{fig: rvsini}). Even though a clear trend is visible in the RV information content (red line), the steep slope of the RV power law (blue line) suggests a genuine dependence on stellar activity.
 
\subsection{$R_{\mathrm{HK}}^\prime$ and color index}
\begin{figure} 
\resizebox{\hsize}{!}{\includegraphics[width=0.5\textwidth]{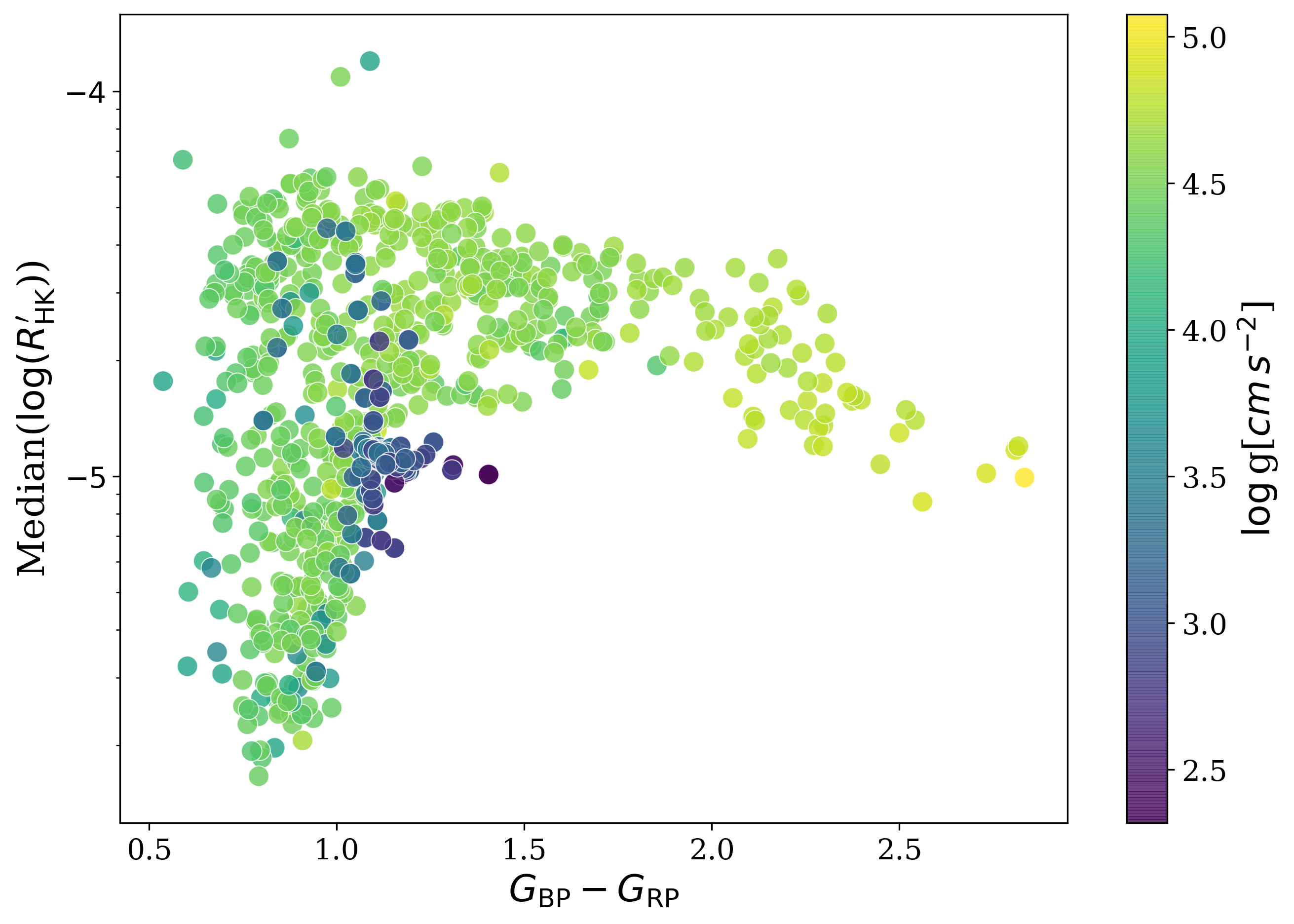}}
\caption{\label{fig: bprplogg} $G_{BP} - G_{RP}$ vs. the median of $\log(R_{\mathrm{HK}}^\prime)$ per star, with logarithmic surface gravity represented as a color map for the clean sample. Only $R_{\mathrm{HK}}^\prime$ measurements that are $>3\sigma$ away from 0 are included.}
\end{figure}

In Fig. \ref{fig: bprplogg} we present, for the clean sample, a plot of $R_{\mathrm{HK}}^\prime$ versus \textit{Gaia}'s photometric color index $G_{BP}-G_{RP}$. Our results do not exhibit a clear Vaughan-Preston gap, a feature first identified by \cite{1980PASP...92..385V}. Such a gap would have suggested a bimodal distribution of the stellar chromospheric activity;  \cite{1980PASP...92..385V} observed that stars predominantly exhibit high or low activity levels in the chromospheric Ca~{\sc ii}~H\&K lines, with a noticeable scarcity of stars displaying intermediate activity levels in the flux-color plot. This apparent gap was considered to stem from different physical causes, such as a statistical bias due to the limited sample size of \citet{1980PASP...92..385V}, a discontinuity in dynamo mechanisms \citep{1981PASP...93..537D}, a rapid rotation spin-down rate \citep{2009A&A...499L...9P}, as well as several other possible origins detailed in \citet{1984ApJ...279..763N}. However, more recent studies, including this work, challenge the strict bimodality of this gap. For example, \citet{2018A&A...616A.108B} found that a significant fraction of stars exhibit intermediate chromospheric activity levels, particularly of $\log\,R_{\mathrm{HK}}^\prime\sim -4.75$, suggesting no apparent lack of intermediately active F and G stars. Their study also identified stars with activity cycle periods that place them between the active and inactive branches, implying that the Vaughan-Preston gap is less distinct than previously assumed. Similarly, our findings do not confirm the presence of an abrupt spin-down at the intermediate activity level.

\begin{figure*}
\centering
\includegraphics[width=0.33\textwidth]{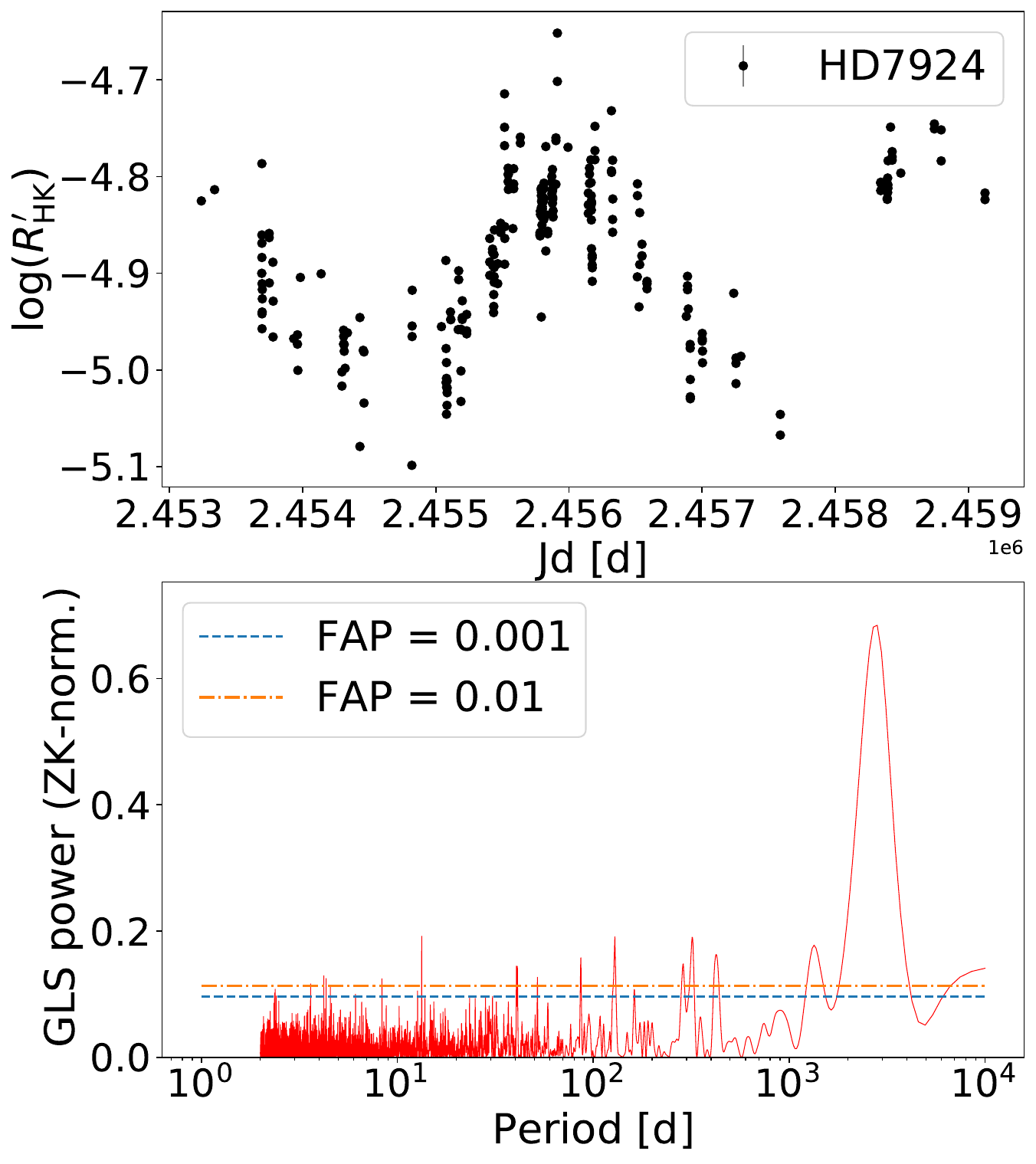} 
\includegraphics[width=0.33\textwidth]{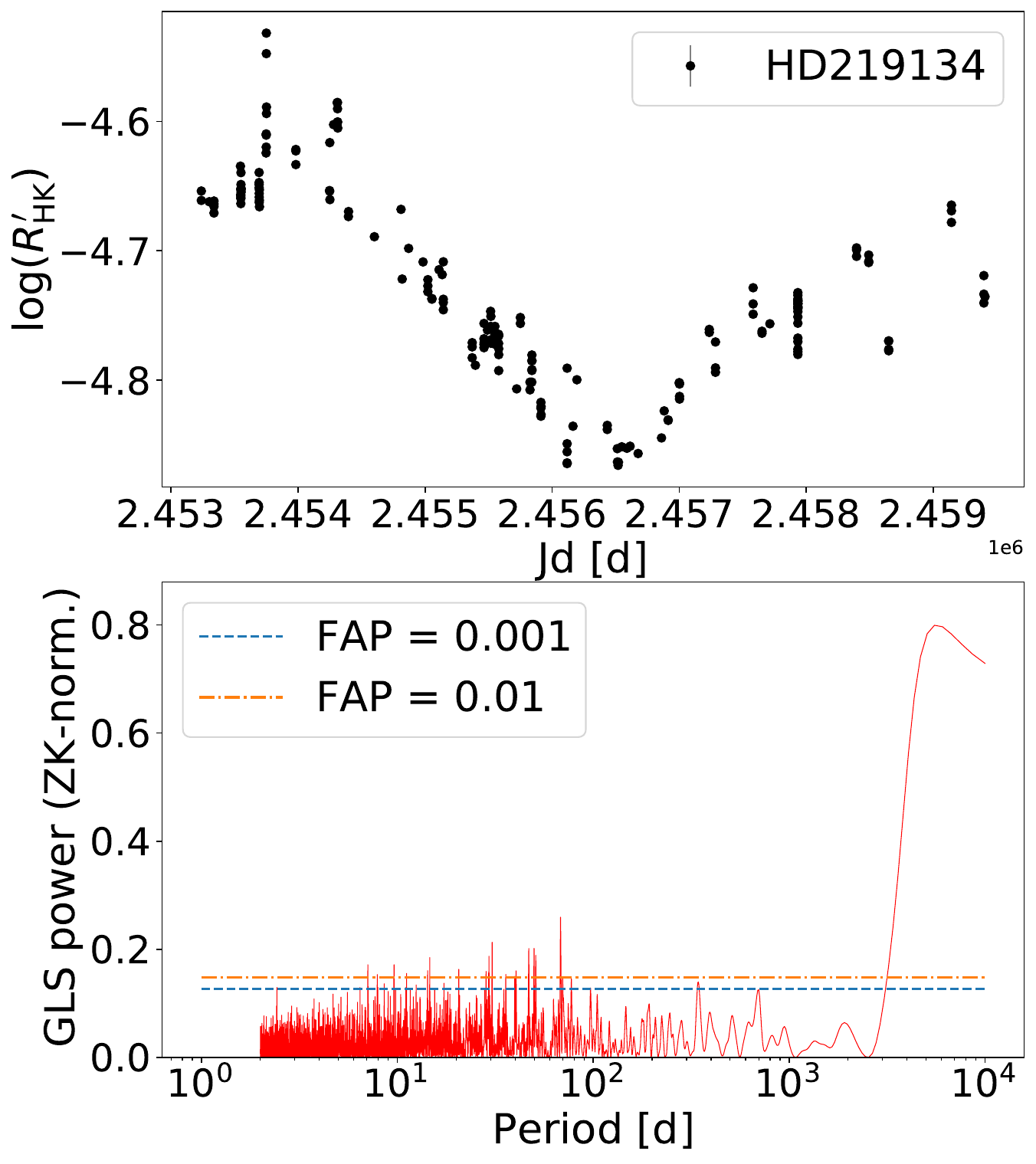}   
\includegraphics[width=0.33\textwidth]{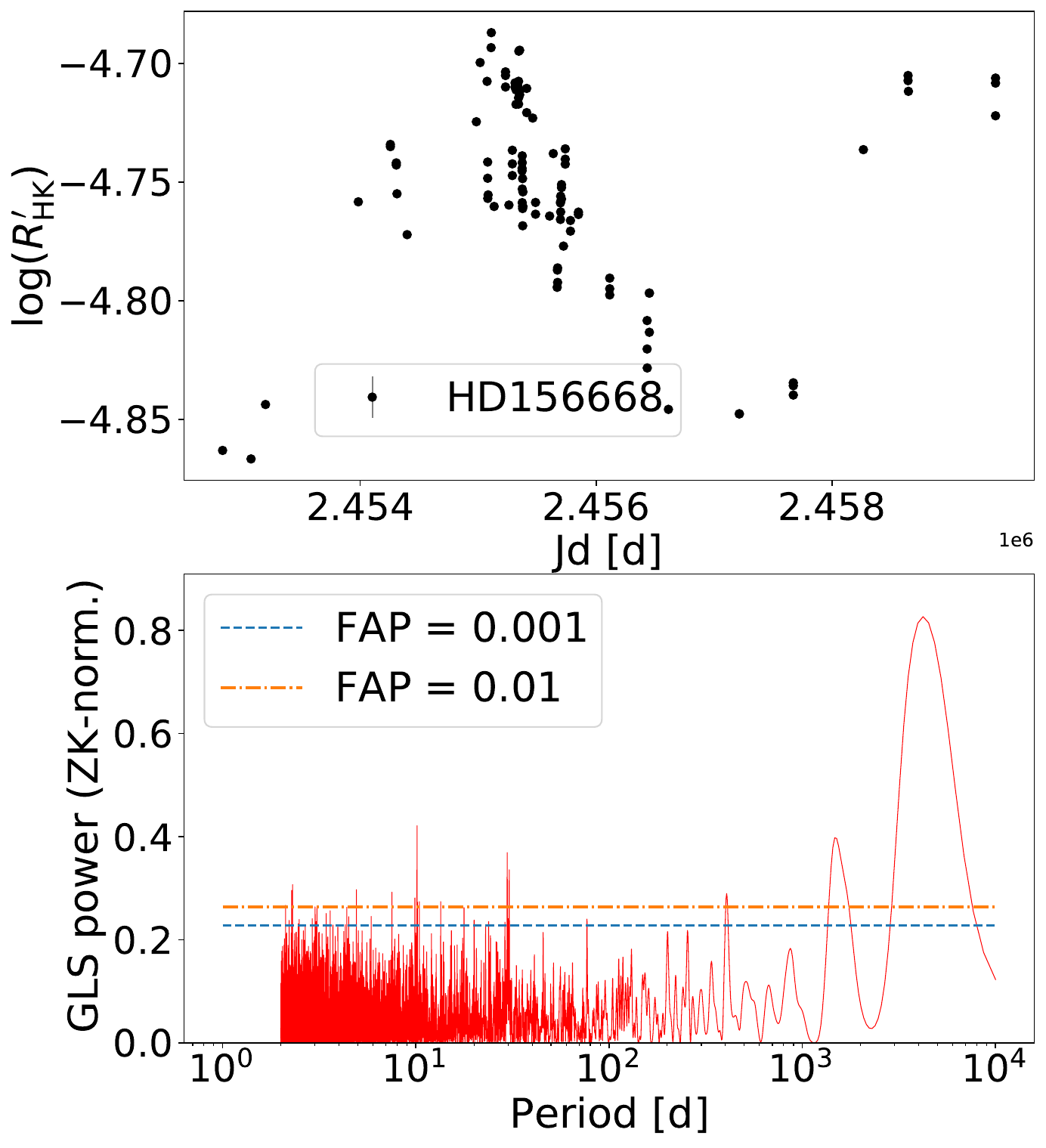}  
\caption{\label{fig: examples} Example of the $R_{\mathrm{HK}}^\prime$ time series (black dots in the upper panel) and the corresponding GLS power spectrum (red line in the lower panel), for HD\,7924, HD\,219134, and HD\,156668. Prominent periodicities in the $R_{\mathrm{HK}}^\prime$ data appear as peaks in the GLS plot. The $1\%$ and $0.1\%$ false-alarm probability levels are indicated in blue and orange, respectively.}    
\end{figure*}

\subsection{Examples} 
\label{sec:compare} 

As an application of our $R_{\mathrm{HK}}^\prime$ measurements, in Fig. \ref{fig: examples} we present the detection of periodic variability, likely caused by solar-like activity cycles, for three stars. The chromospheric emission in Ca~{\sc ii}~H\&K serves as a key indicator for determining both activity cycles and rotation periods, provided that long-term observational data are available \citep{1995ApJ...438..269B, 2018A&A...616A.108B}. To ensure a well-characterized dataset for detecting RV variability driven by stellar activity, we selected the top three stars based on two criteria: more than 30 RV observations and $R_{\mathrm{HK}}^\prime > 0$. The selected stars are HD\,7924, HD\,219134, and HD\,156668. Figure \ref{fig: examples} illustrates the time series of $R_{\mathrm{HK}}^\prime$ for each star, along with the generalized Lomb-Scargle (GLS) periodogram \citep[][]{2009A&A...496..577Z} showing the detected periodic signals. The GLS periodogram is a powerful tool for identifying periodic signals in unevenly spaced data, making it particularly useful in stellar activity studies.

For each of the selected stars: HD\,7924, HD\,219134, and HD\,156668, we determined potential cycle periods of 2817, 5522, and 4214 days, respectively. These values closely match those reported in Table 3 of B17, where they were measured from the S index time series. To evaluate the statistical significance of the detected periods, we incorporated false-alarm probability thresholds of $1\%$ and $0.1\%$ in the GLS periodogram plot. These thresholds estimate the likelihood that the detected signal is not due to random noise. Furthermore, we present the statistical data for these stars in \autoref{table: 3derista}, which summarizes the parameters relevant to the measured stars.

\section{Summary and conclusions} 
\label{sec: summary}

We present a HIRES/Keck precision RV catalog update containing 78,920 RV measurements, incorporating NZP-corrected RVs, for 1,702 stars. As part of this update, measurements of the $R_{\mathrm{HK}}^\prime$, a key indicator of stellar activity, have been added, and problematic spectra for which this measurement was not possible were successfully identified and flagged.

The incorporation of $R_{\mathrm{HK}}^\prime$ measurements is essential for validating exoplanet candidates identified through the RV method, particularly for main-sequence stars, as it provides a key diagnostic for disentangling stellar activity-induced signals from planetary-induced Doppler shift periods and activity cycles. This is particularly important in mitigating false positives, as magnetic activity and starspots can introduce RV variations that mimic planetary signals. By accurately characterizing stellar activity, $R_{\mathrm{HK}}^\prime$ measurements increase the confidence of exoplanet detections and improve the accuracy of orbital parameter determinations.

From the relationship between $RV_{\rm std}$ and $v \sin i$, we observed an RV scatter floor of $2$–$3$\,m\,s$^{-1}$, which is close to the expected precision of HIRES. Furthermore, we have confirmed the expected correlation between activity-induced RV variability and projected rotational velocity.   

We also analyzed the variation in stellar activity between different spectral types by examining $R_{\mathrm{HK}}^\prime$ as a function of the color index, with surface gravity indicated by a color scale. Our results show that most of the HIRES targets are relatively inactive stars. However, for those with well-measured $R_{\mathrm{HK}}^\prime$, the time series data can be utilized to determine stellar rotation periods and activity cycles, further refining activity diagnostics and improving the interpretation of RV signals.

\section*{Data availability}
The full versions of \autoref {table: 1param}, which also contains \autoref{table: 3derista}, and \autoref{table: 2rvlist1} are only available in electronic form at the CDS via anonymous ftp to \texttt{cdsarc.u-strasbg.fr} (130.79.128.5) or via \url{http://cdsweb.u-strasbg.fr/cgi-bin/qcat?J/A+A/}.

\begin{acknowledgements}
We thank the referee for the detailed and useful report. The Israel Science Foundation partially supported this research through grant No. 1404/22. The authors thank Melanie Swain for helping download all HIRES spectra. Most data presented herein were obtained at the Keck Observatory, a private 501(c)3 non-profit organization operated as a scientific partnership among the California Institute of Technology, the University of California, and the National Aeronautics and Space Administration. The Observatory was made possible by the generous financial support of the W. M. Keck Foundation. The authors wish to recognize and acknowledge the highly significant cultural role and reverence that the summit of Maunakea has always had within the Native Hawaiian community. We are most fortunate to have the opportunity to conduct observations from this mountain. This research has used the Keck Observatory Archive (KOA), which is operated by the W. M. Keck Observatory and the NASA Exoplanet Science Institute (NExScI), under contract with the National Aeronautics and Space Administration. This work has made use of data from the European Space Agency (ESA) mission
{\it Gaia} (\url{https://www.cosmos.esa.int/gaia}), processed by the {\it Gaia}
Data Processing and Analysis Consortium (DPAC,
\url{https://www.cosmos.esa.int/web/gaia/dpac/consortium}). Funding for the DPAC
has been provided by national institutions, in particular the institutions
participating in the {\it Gaia} Multilateral Agreement. T. T. acknowledges support from the BNSF program "VIHREN-2021" project No. KP-06-DV/5. We extend our deepest gratitude to Heidelberg University for generously awarding Jerusalem Tamirat the esteemed 4EU+ Research Fellowship, providing invaluable support for three months of dedicated research.
\end{acknowledgements}

\bibliographystyle{aasjournal}
\bibliography{aanda}

\onecolumn
\begin{appendix} 
\section{HIRES main catalog}
\small 
\setlength\tabcolsep{4pt}
\begin{longtable}{l l c c c c c c c c c c}
\caption{Main catalog (extract).} 
\label{table: 2rvlist1}
\addtocounter{table}{-1}\\
\hline\hline
Name & Simbad\_ID & Ra & Dec & BJD & FITS\_MJD & RV & e\_RV & RVC & e\_RVC & Corr & e\_Corr  \\\hline
& & [deg] & [deg] & [d] & [d] & [m\,s$^{-1}$] & [m\,s$^{-1}$] & [m\,s$^{-1}$] & [m\,s$^{-1}$] & [m\,s$^{-1}$] & [m\,s$^{-1}$] \\
\hline
\endfirsthead
HD\,7924 & HD\,7924 & 20.4963 & 76.7102 & 2456676.78469 & 56676.284232 & -2.47 & 1.3 & -2.43 & 1.36 & -0.04 & 0.41  \\
HD\,219134 & HD\,219134 & 348.3207 & 57.1684 & 2457002.72118 & 57002.221139 & -2.68 & 1.07 & -1.10 & 1.28 & -1.58 & 0.70 \\
HD\,156668 & HD\,156668 & 259.4187 & 29.2272 & 2455285.02559 & 55284.524365 & 1.18 & 1.34 & 1.63 & 1.38 & -0.45 & 0.34 \\
HD\,185144 & *sigDra & 293.0899 & 69.6611 & 2459828.72354 & 59828.223531 & -2.69 & 0.86 & -1.956 & 0.983 & -0.734 & 0.476 \\
HD\,42618 & HD\,42618 & 93.0023 & 6.7830 & 2455877.01248 & 55876.511972 & -2.58 & 1.28 & -2.538 & 1.318 & -0.042 & 0.312 \\
\hline
\end{longtable}
\tablefoot{The full catalog for this table, along with the following two continuation tables, is available at the CDS.}

\small 
\setlength\tabcolsep{1.5pt}
\centering
\begin{longtable}{c c c c c c c c c c c c c}
\caption{Continued.} 
\addtocounter{table}{-1} \\
\hline\hline
Count & Exp. & S index & H-index & $R_{\mathrm{H}}^\prime$ & $dR_{\mathrm{H}}^\prime$ & $R_{\mathrm{K}}^\prime$ & $dR_{\mathrm{K}}^\prime$ & $R_{\mathrm{HK}}^\prime$ & $dR_{\mathrm{HK}}^\prime$ & ccf\_RV\_H & ccf\_RV\_K & Med\_ccf\_RV \\\hline
& [s] &  &  &  &  &  &  &  &  & [km\,s$^{-1}$] & [km\,s$^{-1}$] & [km\,s$^{-1}$] \\
\hline
\endhead
76589 & 84 & 0.23 & 0.03 & 2.08$ \times 10^{-6} $& 1.37$ \times 10^{-6} $& 8.61$ \times 10^{-6} $& 7.38$ \times 10^{-7} $& 1.07$ \times 10^{-5} $&  1.56$ \times 10^{-6} $ & -22.7 & -22.7 & -23.3\\
52620 & 0 & 0.24 & 0.04 & 4.92$ \times 10^{-6} $& 8.15$ \times 10^{-7} $& 5.60$ \times 10^{-6} $& 7.37$ \times 10^{-7} $& 1.05$ \times 10^{-5} $& 1.10$ \times 10^{-6} $ & -20.88 & -20.88 & -19.08\\
48293 & 206 & 0.24 & 0.04 & 1.76$ \times 10^{-5} $& 2.18$ \times 10^{-5} $& 8.19$ \times 10^{-6} $& 9.22$ \times 10^{-7} $& 2.58$ \times 10^{-5} $& 2.32$ \times 10^{-6} $ & -194.45 & -194.45 & -45.05 \\
57066 & 98 & 0.0 & 0.0 & 3.09E-5 & 3.22E-6 & 1.69E-5 & 1.99E-6 & 4.78E-5 & 3.78E-6 & -123.42 & -101.22 & 25.98 \\
73504 & 92 & 0.1513 & 0.03089 & -3.59E-6 & 5.30E-7 & -1.21E-6 & 4.18E-7 & -4.81E-6 & 6.75E-7 & -53.48 & -52.88 & -54.08 \\
\hline 
\end{longtable} 
\tablefoot{$R_{\mathrm{H}}^\prime$ and $R_{\mathrm{K}}^\prime$ correspond to distinct measurements of the relative chromospheric emission from the Ca~{\sc ii}~H\&K lines, respectively. The parameters ccf\_RV\_H and ccf\_RV\_K represent the CCF RV values for Echelle orders that contain the Ca~{\sc ii}~H\&K  lines, correspondingly, and Med\_ccf\_RV indicates the median CCF RV per star.}

\small 
\setlength\tabcolsep{1.5pt}
\centering
\begin{longtable}{c c c c c c c c c c c c c c c c c}
\caption{Continued.}
\addtocounter{table}{-1}\\
\hline\hline
$v$flag & $m$flag & RVflag  & est\_snr & EqH & ffH & nfH & nwH & ccffH & snrH & EqK & ffK & nfK & nwK & ccffK & snrK & koaid  \\
\hline
\endhead
1.0 & 0.0 & 0.0  & 139.0 & Pass & 0 & 0 & 0 & 0 & 61.61583 & Pass       &1&     0&      0&      0&      125.40 & HI.20140119.24556.fits \\
1.0 & 0.0 & 0.0 & 123.0 & Pass & 0 & 0 & 0 & 0 & 51.43924 & Pass        &0      &0      &0      &0      &42.8452 & HI.20141211.19105.fits  \\
0.0 & 0.0 & 0.0 & 105.0 & Pass & 0 & 1 & 1 & 1 & 42.25063 & Pass        &0      &1      &1      &1      &36.36179 & HI.20100329.45303.fits  \\
1.0 & 0.0 & 0.0 & 125.0 & Pass & 0 & 1 & 0 & 1 & 60.85466 & Pass & 0 & 1 & 1 & 1 & 54.74191 & HI.20220906.19311.54.fits \\
0.0 & 0.0 & 0.0 & 111.0 & Unknown & \ldots & \ldots & \ldots & 0 & 60.20946 & Unknown & \ldots & \ldots & \ldots & 0 & 40.45288 & HI.20111111.44232.fits  \\
\hline            
\end{longtable} 
\tablefoot{ The flags vflag, mflag, and RVflag correspond to $v \sin i$, [M/H], and RV, respectively. Concise abbreviations used in the table for convenience for each Ca II  H and K lines flag: EqH - EXQUAL H (HIRES internal quality flag: pass, fail and unknown), ffH - flat flux H (plateau-like flat flux), nfH - negative flux H (fluxes below zero), nwH - negative wavelength H (folded wavelengths with negative values), ccfH - cross-correlation function H (difference between the CCF RV we derived and the catalog RV exceeds $3$\,m\,s$^{-1}$), snrH - snr per median column H (from fits header), and the same for K line.}

\newpage
\section{HIRES single-planet systems}
\tiny
\renewcommand{\arraystretch}{0.90}
\begin{longtable}{lccccccc}
\caption{Confirmed single massive planets.} \label{table:appendi} \\
\hline\hline
Planet & $\boldsymbol{\Delta\log\mathcal{L}}$ & $RV_{\mathrm{rms}}$ & Med\_e\_RV & Med\_e\_RVC & K & e\_K & References \\
\hline
 & & [m\,s$^{-1}$] & [m\,s$^{-1}$] & [m\,s$^{-1}$] & [m\,s$^{-1}$] & [m\,s$^{-1}$] & \\
\hline
\endfirsthead

\caption[]{Continued.} \\
\hline
Planet & $\boldsymbol{\Delta\log\mathcal{L}}$ & $RV_{\mathrm{rms}}$ & Med\_e\_RV & Med\_e\_RVC & K & e\_K & References \\\hline

& & [m\,s$^{-1}$] & [m\,s$^{-1}$] & [m\,s$^{-1}$] & [m\,s$^{-1}$] & [m\,s$^{-1}$] &  \\
\hline
\endhead

\hline
\endfoot

\hline
\endlastfoot
HD 210277b & 10.3766 & 2.9623 & 1.12 & 1.176 & 38.32 & 0.255 & 1 \\
HD 16175b & 8.6752 & 4.5107 & 1.505 & 1.553 & 94.0 & 11.0 & 2, 3 \\
HD 45350b & 5.1638 & 3.6572 & 1.2 & 1.25 & 59.4 & 2.9 & 1 \\
HD 222582b & 3.7351 & 3.3887 & 1.385 & 1.4375 & 303.0 & 2.95 & 1 \\
HD 117207b & 3.1763 & 3.371 & 1.505 & 1.5705 & 27.44 & 0.715 & 1, 3 \\
HD 13931b & 2.3411 & 4.0831 & 1.44 & 1.485 & 23.38 & 0.615 & 1, 3 \\
HD 175541b & 2.2376 & 6.1661 & 1.29 & 1.363 & 13.76 & 0.735 & 1 \\
HD 209458b & 2.081 & 11.0145 & 1.75 & 1.794 & 84.9 & 1.25 & 1, 4 \\
HD 190007b & 2.072 & 5.593 & 1.56 & 1.605 & 4.91 & 0.45 & 5 \\
HD 114729b & 2.0463 & 3.9543 & 1.6 & 1.66 & 18.1 & 1.0 & 1 \\
HD 218566b & 1.8617 & 3.7726 & 1.275 & 1.319 & 7.68 & 0.7 & 1 \\
HD 10697b & 1.7392 & 5.9151 & 1.49 & 1.545 & 114.583 & 1.1315 & 3, 6 \\
HD 50554b & 1.5839 & 5.2664 & 1.72 & 1.7605 & 90.8 & 3.35 & 1, 3 \\
HD 86081b & 1.404 & 5.1901 & 1.7 & 1.719 & 205.53 & 0.78 & 7 \\
HD 141937b & 1.112 & 6.2162 & 1.835 & 1.8575 & 234.5 & 6.4 & 2, 8 \\
HD 8574b & 1.0349 & 6.7892 & 1.915 & 1.9585 & 58.3 & 1.55 & 1 \\
HD 142245b & 0.8778 & 5.8723 & 1.195 & 1.2125 & 24.8 & 2.6 & 2, 9 \\
HD 88133b & 0.8754 & 4.0489 & 1.425 & 1.466 & 32.93 & 0.73 & 10 \\
HD 210702b & 0.7562 & 3.7575 & 1.125 & 1.161 & 36.11 & 1.365 & 11 \\
HD 224693b & 0.7561 & 5.5276 & 2.34 & 2.366 & 39.96 & 0.68 & 7 \\
HD 181234b & 0.7045 & 2.8557 & 1.44 & 1.487 & 126.925 & 1.566 & 3, 6 \\
HD 38801b & 0.6873 & 8.7225 & 1.175 & 1.222 & 196.3 & 3.8 & 10 \\
HD 189733b & 0.6519 & 15.0117 & 1.04 & 1.069 & 204.7 & 2.55 & 4, 12 \\
HD 109749b & 0.5158 & 3.2765 & 1.34 & 1.371 & 29.2 & 1.1 & 7 \\
HD 1502b & 0.5012 & 12.7243 & 1.55 & 1.568 & 57.5 & 3.3 & 10 \\
HD 179949b & 0.467 & 7.8666 & 2.28 & 2.291 & 118.0 & 2.8 & 1 \\
HD 52265b & 0.4165 & 4.42201 & 1.59 & 1.62 & 41.14 & 0.69 & 1 \\
HD 131496b & 0.4111 & 6.5044 & 1.23 & 1.257 & 31.6 & 1.8 & 10 \\
HD 94834b & 0.3867 & 5.9949 & 1.27 & 1.299 & 20.7 & 2.9 & 10 \\
HD 102956b & 0.2596 & 6.6496 & 1.245 & 1.289 & 74.6 & 1.8 & 10 \\
HD 10442b & 0.2175 & 5.9408 & 1.8 & 1.8185 & 29.9 & 1.6 & 10 \\
HD 72490b & 0.2002 & 6.0459 & 1.45 & 1.5095 & 33.5 & 1.5 & 10 \\
HD 152581b & 0.1218 & 4.1966 & 1.44 & 1.449 & 36.2 & 1.3 & 10 \\
HD 211810b & 0.1055 & 3.5015 & 1.28 & 1.317 & 15.6 & 7.2 & 7 \\
HD 130322b & 0.0708 & 6.1316 & 1.42 & 1.473 & 111.1 & 2.35 & 1 \\
HD 192263b & 0.0341 & 8.2431 & 1.47 & 1.548 & 55.3 & 2.15 & 12 \\
HD 231701b & -7.0E-4 & 6.9598 & 2.405 & 2.443 & 39.2 & 1.2 & 7 \\
HD 17156b & -0.0155 & 3.5644 & 1.46 & 1.494 & 274.7 & 3.5 & 4, 7 \\
HD 16417b & -0.0443 & 2.6798 & 1.045 & 1.064 & 5.0 & 0.4 & 13 \\
HD 40979b & -0.0555 & 18.2648 & 2.21 & 2.247 & 107.9 & 4.0 & 1 \\
HD 190228b & -0.112 & 3.9622 & 1.715 & 1.7415 & 92.936 & 1.2395 & 3, 6 \\
HD 31253b & -0.1181 & 5.3455 & 1.555 & 1.596 & 10.8 & 1.8 & 1 \\
HD 170469b & -0.1295 & 2.6718 & 1.52 & 1.577 & 10.2 & 1.3 & 1 \\10
HD 168746b & -0.1305 & 3.5392 & 1.635 & 1.662 & 27.17 & 0.789 & 1 \\
HD 4313b & -0.151 & 5.6019 & 1.415 & 1.4515 & 40.3 & 1.7 & 10 \\
HD 206610b & -0.3093 & 10.2176 & 1.35 & 1.3765 & 35.4 & 1.0 & 10 \\
HD 99109b & -0.3885 & 5.8102 & 1.42 & 1.477 & 12.98 & 0.875 & 1 \\
HD 96167b & -0.5356 & 4.3132 & 1.45 & 1.504 & 21.5 & 1.4 & 10 \\
HD 214823b & -0.5451 & 3.5182 & 1.775 & 1.8005 & 283.005 & 0.7795 & 3, 6 \\
HD 164509b & -0.5612 & 5.4072 & 1.285 & 1.317 & 13.15 & 0.77 & 14 \\
HD 136925b & -0.5987 & 5.2447 & 1.645 & 1.6825 & 11.4 & 0.995 & 1 \\
HD 108863b & -0.6751 & 6.1651 & 1.2 & 1.243 & 47.4 & 1.5 & 10 \\
HD 178911b & -0.7386 & 3.9829 & 1.28 & 1.315 & 342.69 & 0.82 & 1 \\
HD 55696b & -0.7716 & 8.9351 & 2.06 & 2.085 & 76.7 & 3.9 & 3, 7 \\
HD 149026b & -0.8982 & 7.0589 & 1.52 & 1.547 & 39.22 & 0.68 & 4, 7 \\
HD 82886b & -0.9558 & 8.2932 & 1.62 & 1.679 & 28.7 & 2.1 & 2, 9 \\
BD 103166b & -0.983 & 6.4099 & 2.06 & 2.092 & 60.2 & 1.5 & 1 \\
HD 98219b & -1.2203 & 5.8728 & 1.29 & 1.34 & 42.0 & 2.1 & 10 \\
HD 154345b & -1.2958 & 4.14601 & 1.45 & 1.4825 & 13.29 & 1.125 & 1, 3 \\
HD 149143b & -1.4233 & 7.4232 & 2.18 & 2.198 & 150.0 & 0.65 & 7 \\
HD 185269b & -1.6229 & 6.5588 & 2.08 & 2.0945 & 93.3 & 1.4 & 10 \\
HD 102195b & -1.8258 & 8.4974 & 1.17 & 1.203 & 66.9 & 3.4 & 12 \\
HD 43691b & -1.9087 & 6.27601 & 1.72 & 1.748 & 130.06 & 0.84 & 7 \\
HD 32963b & -2.0956 & 2.9379 & 1.355 & 1.3845 & 11.32 & 0.415 & 1, 3 \\
HD 73534b & -3.8289 & 4.6557 & 1.23 & 1.272 & 17.11 & 0.93 & 10 \\
HD 179079b & -4.3308 & 4.0937 & 1.41 & 1.434 & 6.22 & 0.78 & 7 \\
HD 49674b & -4.3493 & 5.8716 & 1.22 & 1.28 & 13.21 & 0.885 & 1 \\
HD 180617b & -4.4358 & 5.7181 & 2.62 & 2.632 & 2.696 & 0.224 & 15 \\
HD 87883b & -5.7473 & 2.3032 & 1.36 & 1.423 & 51.435 & 0.626 & 6 \\
HD 46375b & -7.0105 & 3.5297 & 1.21 & 1.293 & 33.46 & 0.8 & 1 \\
HD 45652b & -9.5988 & 14.3901 & 1.06 & 1.162 & 33.2 & 1.8 & 7 \\

\end{longtable}
\tablebib{
(1)~\citet{2021ApJSRosenthal}; (2) \citet{2017AJStassun}; (3) \citet{2023RAAXiao}; (4) \citet{2023ApJSKokoriK};
(5) \citet{2023AStalportS}; (6) \citet{2022ApJSFengF}; (7) \citet{2018AJMentM}; (8) \citet{2002AUdryU};
(9) \citet{2011ApJSJohnsonJ}; (10) \citet{2019AJLuhn}; (11) \citet{2023PASJTengT}; (12) \citet{2021AJParedesP}; (13) \citet{2009ApJTooleO}; (14) \citet{Giguere2011}; (15) \citet{2021AJBurtB}.
}
\tablefoot{ $RV_{\mathrm{rms}}$ is the root mean square of the corrected RVs. Med\_e\_RV and Med\_e\_RVC represent the median uncertainties of the uncorrected and corrected RVs, respectively. K is the semi-amplitude of the Keplerian signal, with its associated uncertainty denoted as e\_K.}
\end{appendix}
\end{document}